\documentclass[namedreferences,hyperref,optionalrh]{spr-sola}
\usepackage{graphicx}        
\usepackage{color}           
\usepackage{lmodern}

\chardef\us=`\_

\begin{document}

\begin{frontmatter}
\title{Structures of various
types of symmetry in the solar activity cycle}

\author[addressref={aff1},email={e-mail. obridko@mail.ru}]{\inits{V.N.}\fnm{V.N.}~\snm{Obridko}\orcid{ 0000-0001-5100-806X}}
\author[addressref=aff1, email={e-mail.as.shibalova@physics.msu.ru}]{\inits{A.S.}\fnm{A.S.}~\snm{Shibalova}\orcid{0000-0002-7228-6226}}
\author[addressref={aff1,aff2}, corref, email={e-mail.sokoloff.dd@gmail.com}]{\inits{D.D.}\fnm{D.D.}~\snm{Sokoloff}\orcid{0000-0002-3441-0863}}
\address[id=aff1]{IZMIRAN, 4, Kaluzhskoe Shosse, Troitsk, Moscow, 108840, Russia}
\address[id=aff2]{Department of Physics, Lomonosov Moscow State University, Moscow, 119991, Russia and Moscow Center for Fundamental and Applied Mathematics, Moscow, 119991, Russia}

\runningauthor{Obridko et al.}
\runningtitle{Structure and tesseral harmonics in solar cycle}

\begin{abstract}
The solar cycle is a complex phenomenon. To comprehensively understand it, we have to study various tracers. The most important component of this complex is the solar dynamo, which is understood as self-excitation of the solar magnetic field in the form of traveling waves somewhere in the convection zone. Along with the solar dynamo, the formation of the solar cycle involves other processes that are associated with the dynamo but are not its necessary part. We give a review of such phenomena that have not yet been explained in terms of dynamo theory. We consider the manifestations of the solar cycle in harmonics of the solar large-scale surface magnetic field, including zonal, sectorial, and tesseral harmonics; analyze their contribution to magnetic energy; and identify phases of the activity cycle using harmonics of different types of symmetry. The universal magnetic scenario of a solar activity cycle does not depend on its number and height. At the beginning of the cycle on the photosphere, the zonal harmonics account for 37-42\% of the total energy (not 100\%, as assumed in simplified descriptions). Sectorial harmonics do not disappear at all but account for 5-10\% of the total energy. At this stage, the greatest energy (about 40\%) is contained in the tesseral harmonics. As the cycle develops, the relative energy of zonal harmonics gradually decreases, reaching a minimum of 15-18\% immediately before the onset of the sunspot maximum. The relative energy of sectorial harmonics increases and reaches a maximum (60-65\%) somewhat later than the calendar date of the sunspot maximum. A particular feature of the tesseral harmonics is that their relative energy index changes in a much narrower range and never falls below 40\% even at the cycle minimum. This is due to active regions and nonglobal magnetic fields. It is possible that tesseral harmonics are formed in shallow subphotospheric layers.
\end{abstract}
\end{frontmatter}

\section{Introduction}
     \label{S-Introduction} 
     
The main present-day concept of the origin of solar activity is based on mean-field dynamo theory (e.g.  see \cite{CG95,CS23, BS05, R19}, about more complicated ideas like small-scale dynamo see, e.g. \cite{XL21, Retal23}). The differential rotation transforms the initial poloidal magnetic field into a toroidal one. Then, the meridional circulation and/or the Babcock-Leighton mechanism (possibly with the participation of the Parker mechanism) restore the poloidal magnetic field near the poles.  The propagation velocity of the dynamo wave is controlled by the meridional circulation. This is how an 11-year cycle of magnetic energy variations arises. The connection between the two successive cycles is achieved by the fact that the restored polar magnetic field, which is the beginning of the following cycle, is opposite in sign to the field in the previous cycle. As a result, the sign of the toroidal magnetic field in cycle $n$ is opposite to that in cycle $n-1$. This is manifested as Hale's polarity law.	

This mechanism alone, without additional assumptions, does not explain why the cycles differ in height. (Of course, various mechanisms for this are discussed in the above-mentioned papers.) 

In addition, the dynamo (the so-called mean-field dynamo) accounts mainly for the large-scale field. How spots and active regions arise from this field is not quite clear, either. The basic mean-field dynamo models cannot directly describe elements of such characteristic scales as sunspots. Apparently, the mean-field dynamo generates a 'toroidal' magnetic flow near the solar surface. The mechanism of transformation of this field into a sunspot field is not yet clear. This could be the “effect of negative magnetic pressure” \citep{Kleetal89, Brandetal2010} or the heat instability proposed by \cite{Kitchma2000}. At the same time, the question of self-similarity of the internal structure of cycles has not been studied.We do not know whether the different components of the cycle depend only on its height or not. It turns out that the relative contribution of harmonics of different types to the magnetic energy is virtually independent of the cycle height \citep{obretal24}.

Some progress in predicting the height of the cycle a few years prior to the maximum has been made using some precursors, including polar-field data \citep{obrshel09, obrshel17, obretal23a}. (See also a comprehensive review by Dibyendu Nandy \citep{nandy21}). The available data indicate a high correlation between the magnitude of the polar field and the number of sunspots in the following cycle \citep{Biswasetal23}.

Recently, works have appeared devoted to the forecast of the polar field several years before the onset of the minimum. That is, the amplitude of the following cycle can be predicted several years after the change in the polar field in the previous cycle, which significantly increases the forecast lead time \citep{Pishvas23, Kumaretal21, Kumaretal22}.

Finally, the most promising method is direct use of the dynamo mechanism with appropriately selected parameters. The difficulty here is the choice of optimal parameters, of which there are so many that their direct selection is impossible. In addition to that, the question of how one cycle differs from another has not yet been finally resolved. In fact, we have to select the contribution of the stochastic component for each cycle separately. We do not know to what extent the solar dynamo is determined by stochastic or deterministic processes \citep{Minetal2002, Minetal2004, Kit2016, Kitkos08, Kitkos11}.

Direct data methods (data assimilation, for example \citep{KK11}) have been used to solve this problem. These methods combine observational data and models to efficiently and accurately estimate the physical properties that cannot be observed directly.

As a rule, the physically based forecasting methods calculate the polar field as a precursor of the following cycle. These forecasts use flux-transport-dynamo models (FTD), surface-flux-transport models (SFT), or a combination of both \citep{Upthat18, Jiangetal18, Bhownan18, Labetal19, Jiangetal23, Bhowetal23, Guoetal21}.

In the above context, the aim of this paper is to consider not only the simplest harmonics, but also the harmonics of the solar large-scale surface magnetic field of various types of symmetry, including tesseral ones. 

\section{Initial data and analysis methods}

Here, we focus our analysis on the structure of the large-scale global magnetic field as a combination of multipoles of different orders. We are interested in the time evolution of individual multipoles as well as their correlations.

The amplitudes and phases of lower-order multipoles and their correlations with the solar photospheric magnetic field have been studied in various papers based on Kitt Peak and WSO data (for example, \citep{Lev77, Hoeks84, Hoeks91, Stenvog86, Stenwei87, Goketal92a, Goketal92b, Knasten2005}.

We will use the spherical analysis method to estimate the contribution of various structural parameters to the cyclic evolution of large-scale fields. The analysis is based on Stanford photospheric magnetic data with a resolution of 3 arc min for the period of June 1976 to July 2022 (i.e., Carrington rotations  1642—2055)  obtained from http://wso.stanford.edu/forms/prsyn.html. It is assumed that the magnetic field  in a spherical layer  from the photosphere ($R_0$) to a fixed spherical surface of radius $R_s$, conventionally called the source surface is fully described by the potential approximation (potential-field source-surface model, PFSS) \citep{Hoeks84, Hoeks91}. This yields the set of magnetic field components $B_r$, $B_\theta$, and $B_\phi$ in the spherical coordinates $r$, $\theta$, and  $\phi$ (radial, latitudinal, and longitudinal, respectively). In this work we will use only the following expression for $B_r$:

 \begin{eqnarray}
 B_r=\sum_{l,m}P_l^m (\cos \theta) (g_i^m \cos \phi + h_i^m \sin m \phi)\times\\
 \nonumber
 \times ((l+1) (R_0/r)^{l+1} - l(r/R_s)^{l+1}c_l
 \end{eqnarray}
 Here, $0 \le m$, $l < N$ (usually, $N \le 9$), $c_l=-(R_0/R_s)^{l+2}$, $P_l^m$  are the Legendre polynomials, and $g_l^m$, $h_l^m$  are the harmonic coefficients.
The latter are calculated from WSO Stanford data. To find the harmonic coefficients, $g_l^m$ and $h_l^m$, and thus, to fully determine the solution, we had to use boundary conditions. The radial component of the magnetic field  on the photosphere surface plays the role of one boundary condition, while the other is the assumption  that the magnetic field at the source surface is purely radial.

In order to estimate the magnetic energy at a sphere of radius $r$, we introduce as an index the mean value of $B_r^2$ taken at the sphere of radius $r$ \citep{Oye89,Setal89, Oye92}:

\begin{equation}
i(B_r)_{|r} = <B_r^2>_{|r}\,,
\end{equation}

Large-scale field structures can be separated using harmonic coefficients. This is physically important because, as shown in \citep{obretal24}, the three types of symmetry naturally change each other during the cycle and often reflect different types of solar activity, both in spatial dimensions and in forms of activity.
 
The large-scale magnetic fields can be divided by harmonic coefficients into the following structures of different types of symmetry \citep{Oye89, Setal89, Oye92}:
\begin{itemize}
 \item Zonal structures ($m=0$). Harmonics are called zonal if their values change only with latitude. Within one zone limited by parallels from the north and south, the zonal harmonic retains its value and sign. The zonal harmonics, in turn, can be divided into zonal odd -- ZO ($l$ is odd, the most important case is the axial dipole, $l$=1) and zonal even --
ZE ($l$ is even, the most important case is the axial quadrupole, $l$=2);
 \item Sectorial structures $m=l$. Harmonics are called sectorial if their values change only with longitude. Within one sector, limited by meridians, the sectorial harmonic retains its sign. These, in turn, can be divided into sectorial odd -- SO ($l$ is odd, the most important cases are the equatorial dipole and two-sector structure with $l=m=1$) and sectorial even -- SE ($l$ is  even, e.g., $l=m=2$);
 \item Tesseral structures ($m\neq l$); i.e., all other harmonics. These harmonics were named so, because the curves on which they vanish are parallels of latitude, $l-m$, and meridians, $l-m$, dividing the surface of a sphere into rectangles (tessaras)  \citep{Whitwat90} ("tessera" -  a tile of roman mosaic). 
\end{itemize}

Of course, it is possible to introduce partial energy indices for each type of symmetry, the sum of which will be equal to the general indices discussed above. We carried out such calculation and normalized the partial indices to the general ones in order to exclude the general drop of solar activity in the past solar cycles.

The zonal and sectorial harmonics along one of the coordinates (latitude or longitude) cover the entire circumference of the Sun and form either the zones limited in latitude or spherical sectors limited in longitude. Of course, they cannot directly describe local magnetic fields, although they are probably related to them at the generation level. Tesseral harmonics can describe fields of relatively small spatial scales. In principle, they can describe local fields to the accuracy of the original data.

\section{Contribution of different  structural components}

\begin{figure}    
\includegraphics[width=0.85\textwidth]{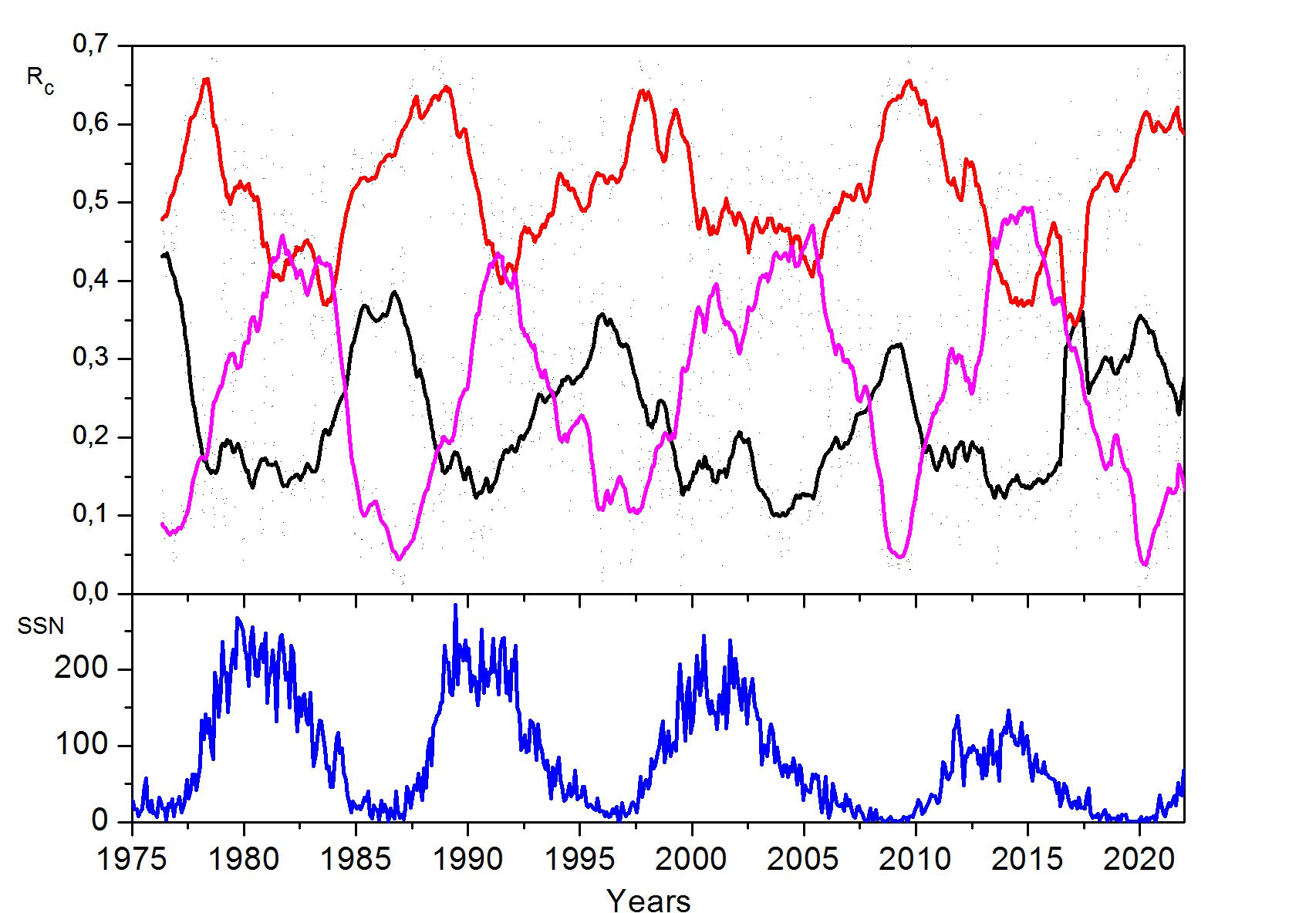}
\small
        \caption{Time dependence of the relative contribution ($R_c$) of three structural components at the photosphere level (upper panel): zonal (black), sectorial (magenta), and tesseral (red). The lower panel shows the monthly mean sunspot numbers SSN.
       }
\label{F1}
\end{figure}

The time dependence of the relative contribution of three structural components at the photosphere level is shown in Fig.~\ref{F1}. The data were calculated for each half of the Carrington rotation and smoothed over 27 points (i.e. at approximately 1-year intervals). The lower panel shows the monthly mean sunspot numbers SSN.

\begin{figure}    
\includegraphics[width=0.85\textwidth]{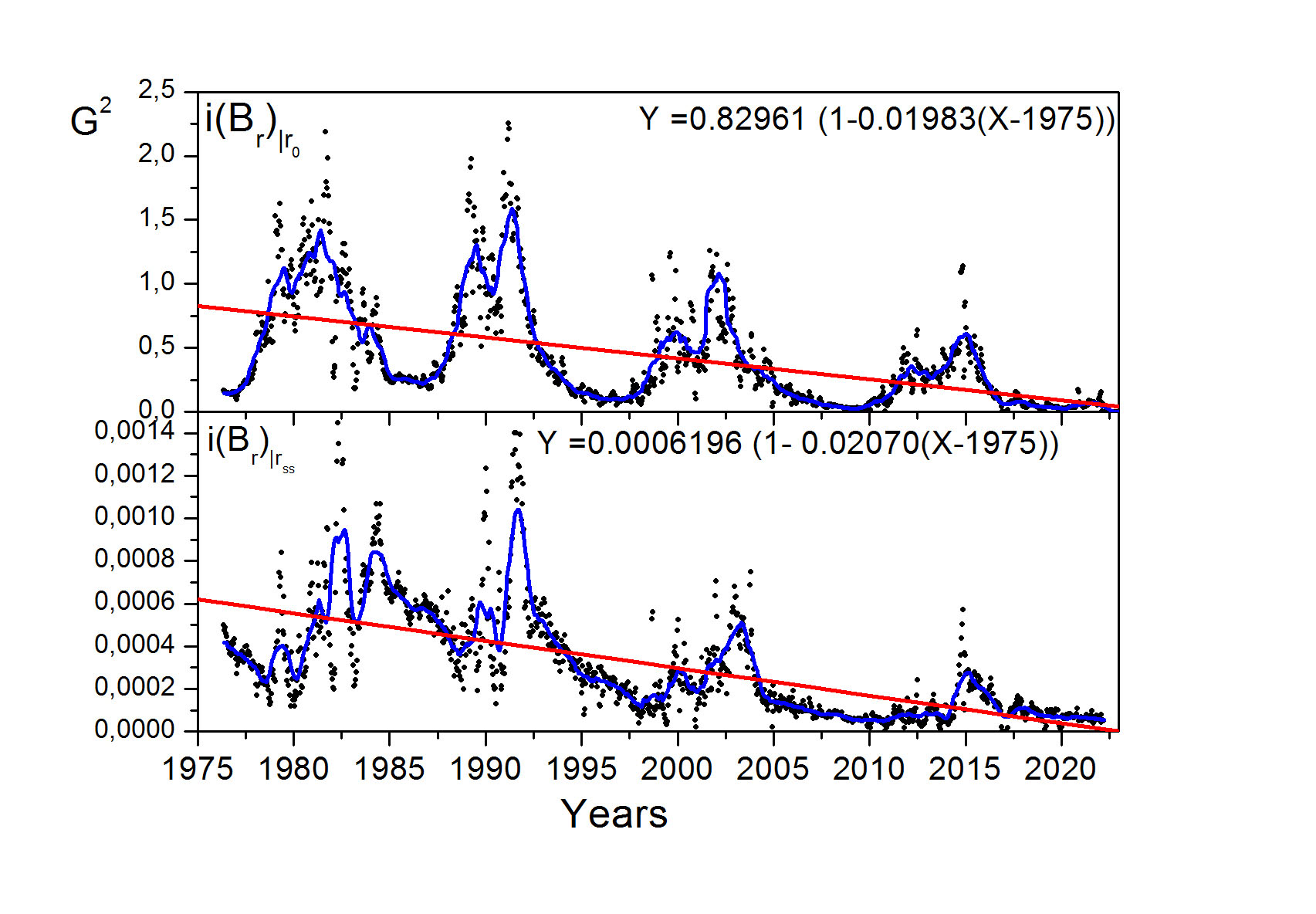}
\small
        \caption{Comparison of the $iB_r$ index on the photosphere (top) and on the source surface (bottom). The dots show data for every half Carrington rotation, the thick blue lines are the result of smoothing over 27 points, and the straight red lines show the linear trend obtained by least-square fitting (the equation in the upper right corner of each panel describes the trend)).
       }
\label{F2}
\end{figure}

The self-similarity is striking. The cycles of all energy indices are similar both in shape and in height. This is all the more surprising because the general level of solar activity was steadily decreasing over the past four cycles. This is readily seen in the sunspot number index (Figure 1, bottom panel) and is also noted in almost all other indices, e.g. \citep{sval13, hath15}. The relationship between the activity indices manifests itself in the Gleissberg cycle, the Gnyshev-Ohl rule, and in the expected period of low cycles. This trend is clearly pronounced in variations of the above-introduced mean square solar magnetic field index $iB_r$, which can be calculated both on the photosphere and above, at the source surface level (2.5 solar radii), see Fig.~\ref{F2}.

The trend for the past 4 cycles is clearly visible in Fig. 2. It is interesting to note that the angular coefficients for the trends on the photosphere and on the source surface almost coincide. This could not be expected in advance, since each value on the plot was obtained as a result of calculations for 100 multipoles, each having its own dependence on the height and characteristic dimensions. The coincidence of all relative contribution ($R_c$) of three structural components indicates that  the contributions of the structural components are associated with the mechanism responsible for the 11-year cycle while the trend is caused by a mechanism independent of the one that generates the 11-year cycle.

Turning back to Figure 1, we note that, although the height of the cycles varies significantly both in the spot numbers and in $iB_r$, the relative proportion of structures (partial indices) persists with amazing accuracy. The stability and actual independence of zonal and sectorial modes that are in antiphase surprise. The tesseral modes never fall to zero, ranging from 0.3 to 0.6 and reaching a maximum in the growth phase.

Thus, in each cycle there is an internal hierarchy that determines the relationship between structures with different types of symmetry. It does not depend on the total energy of a given cycle and is generally consistent with the basic concept of the solar dynamo. The change in the height of the cycles is due to some other modulation mechanism (e.g. \cite{T97, Petal18, J20, Tetal22, Betal23} ). Note, that the observational time series are too short to choose among the mechanisms suggested the most appropriate one.

Now, we can quantitatively describe a universal magnetic scenario of the solar cycle, which does not depend on the cycle height and number. At the beginning of the cycle on the photosphere, the zonal harmonics account for 37-42\% of the total energy (not 100\% as the simplified descriptions assume). The sectorial harmonics do not vanish, but make up 5-10\% of the total energy. At this stage, most of the energy (about 40\%) is contained in tesseral harmonics, which are not associated with either sunspot numbers or global fields. As the cycle develops, the relative energy of the zonal harmonics gradually decreases, reaching a minimum of 15-18\% immediately before the onset of the sunspot maximum. The relative energy of sectorial harmonics increases and reaches a maximum of 60-65\% soon after the calendar sunspot maximum (at the time of the SSN secondary maximum or at the beginning of the decline phase). The tesseral harmonics rise for some time reaching a maximum of 60-65\% before the onset of SSN maximum and then, fall down to their standard value of 40\% and begin to grow again. At the beginning of the SSN decline phase, simultaneously with the minimum of the zonal harmonics, a new growth of sectorial and tesseral components begins.

The fact that the cycles are not equal in height is not a fault of the dynamo mechanism! The classical dynamo generates cycles that are strictly equal up to fractions of the total energy, $iB_r$. Additional considerations are needed to account for unequal cycles. Stochastic factors are often introduced to explain individual cycles. However, this does not explain the long-term modulation on a scale of several cycles. Apparently, there is an external factor causing multi-cycle modulation (the factor may be associated with a Gleissberg cycle or a longer cycle, however, the time series exploited are too short for a definite decision).

The process of generation of the magnetic field is mainly controlled by the dynamo mechanism. This is a fairly precise cause-and-effect deterministic process. Of course, the mechanism admits deterministic chaos and/or statistical fluctuations of dynamo drivers however our database is too short to isolate these options.) It ensures division into three main components (zonal, tesseral, and sectoral) that is stable in time, the same in all cycles, and has a strict dating within the cycle.

\section{Analysis of the contribution of tesseral harmonics}

 We were surprised to learn that the analysis of tesseral harmonics is virtually absent in the  astronomical scientific literature. The search via the ADS abstract service has revealed only two publications on solar physics mentioning tesseral harmonics during the entire existence of this database.

\cite{Ziegetal19} studied the power spectra of multipole moments by separating the global magnetism into the zonal, sectorial, and tesseral harmonic components. However, since during the solar minimum, the axisymmetric component dominates the large-scale photospheric (and coronal) magnetic field, the authors limited their analysis to the zonal harmonic spectrum. 

\cite{Mikh20} analyzed the tesseral component of the quadrupole and discovered a hidden cycle of about half an 11-year interval (72 solar rotations).

This lack of attention to tesseral harmonics is not surprising. First, the zonal and sectorial harmonics have a simple and clear physical meaning. They follow directly from the simplest dynamo models and are easy to describe and interpret. In this study, we limited ourselves to analyzing 9 harmonics. This means that the full set of expansion coefficients is 100. 

We will not take into account the coefficient $g_{00}$, which has no physical meaning and is probably associated with observation errors. The remaining coefficients give us 9 zonal and 9 sectorial harmonics, and the rest are tesseral harmonics. Their analysis is extremely complex, especially because one will have to take into account the dependence not only on time and latitude, but also on longitude. For such an analysis, it would be necessary to work out some special tools for studying and presenting the results. Even simpler problems of cyclic variation and latitude-longitude evolution of an inclined dipole and quadrupole required very non-standard methods of presentation.

Nevertheless, tesseral harmonics cannot be ignored. It is not only because they are the absolute majority of the expansion  harmonics, but because they  account for the main part of energy of the solar surface magnetic field. Indeed, it is clear from Figure 1 that the energy of tesseral harmonics on the photosphere makes up from 40 to 70\% of the total energy of the magnetic field in any phase of the cycle. It is not surprising. We know that the flux of local magnetic regions is comparable to that of large-scale magnetic fields.  However, from the point of view of energy indices, the situation is radically different. The specific energy corresponding to a unit of surface in local fields is six orders of magnitude higher than in global ones. Therefore, the contribution of local fields to the energy index is quite significant.

\begin{figure}    
\includegraphics[width=0.85\textwidth]{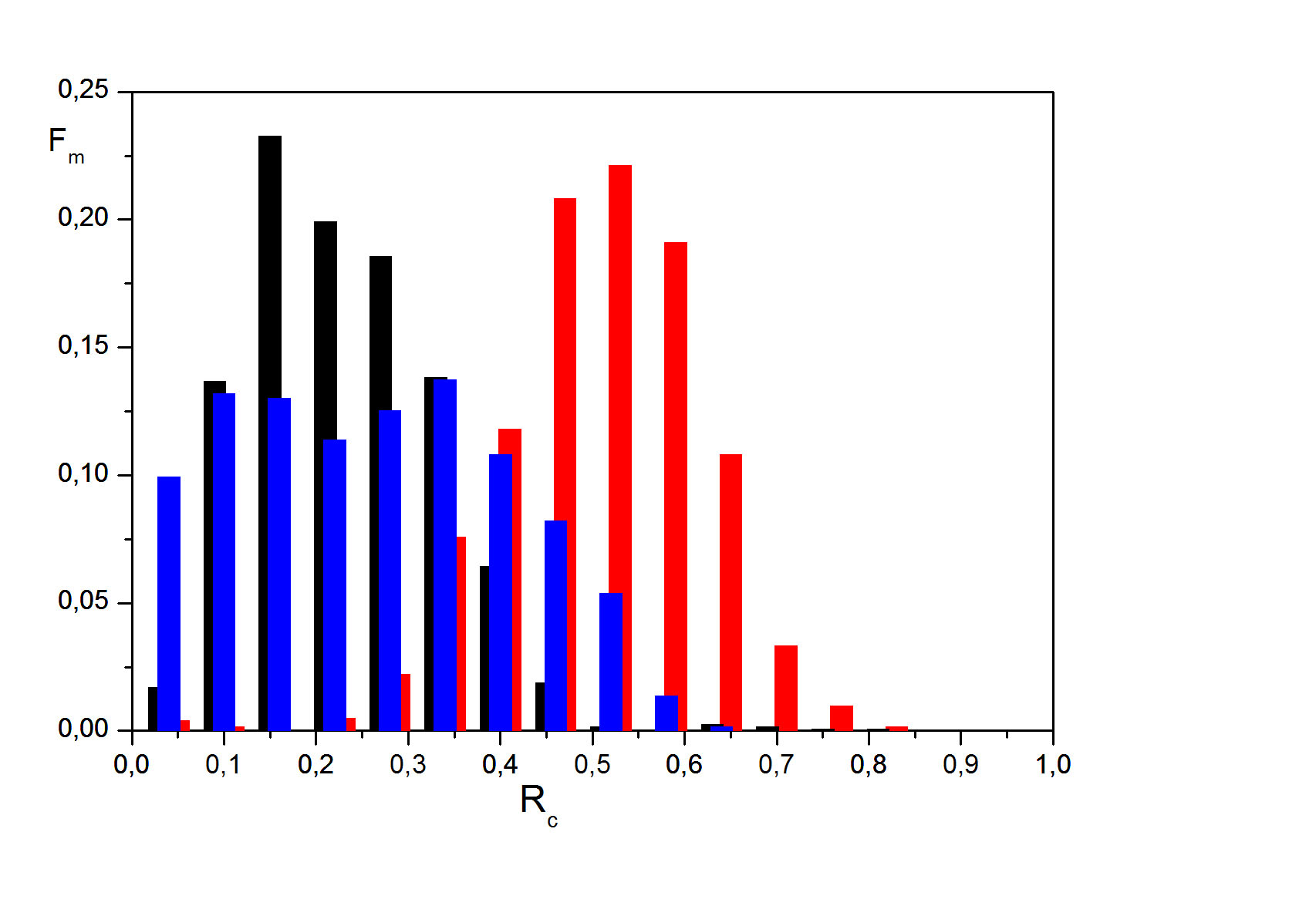}
\small
        \caption{Histograms of 
        distribution of the contribution of fields of three types of symmetry: zonal (black curve), sectorial (magenta), and tesseral (red).
       }
\label{F3}
\end{figure}

A particular feature of tesseral fields is that they  not only exceed the fields of other symmetry types in total energy, but their energy does not drop to zero even at the minimum of the cycle. This may be due to the fact that small elements with a strong magnetic field exist even at the sunspot minimum (see \citep{Obretal23b} and references therein). On the other hand, it is possible that the mean-field dynamo can generate fields of non-global scale that obey the 11-year periodicity, but do not vanish at the time of the minimum.

In order to clarify the relative contributions of magnetic harmonics of various symmetries let us move in Fig.~\ref{F3} from
plotting temporal dependence's to a statistical presentation. In order to visualize the relative contribution of magnetic harmonics of different symmetry, let us pass from time dependencies to statistical presentation. We have plotted the relative number of the solar half-rotations, in which  the magnetic harmonics of a given type of symmetry contributed a given part of energy to the solar surface magnetic field (Fig.~\ref{F3}). 

The histogram in figure~\ref{F3} shows the distribution of fields of three types of symmetry. A total of 1229 solar half-rotations took place for 1976-2022. In this period, the relative contribution of the zonal structures (black columns) ranged from 0 to 0.4, and its maximum value and total duration fell on the epochs of sunspot minima. The sectorial fields (magenta) showed a much wider distribution. Their relative contribution ranged from 0 to 0.6, and they were observed not only at the maximum of the cycle, but also during a significant part of the ascending and descending branches. Conversely, the distribution of the tesseral harmonics (red columns) was much narrower. They were generally more powerful than the other types of symmetry and their relative contribution ranged from 0.3 to 0.65. The diagram in figure~\ref{F3} based mainly on data from figure~\ref{F1}, shows the occurrence rate of three different modes $F_m$ (zonal, sectorial, and tessaral) as a function of the relative contribution of each one, $R_c$. In general, the contribution of the tesseral component exceeds the total contribution of all the others (i.e., $R_c>0.5$) during 78\% of the entire period under consideration (i.e., about 36 years).

For the above reasons, we do not try to analyze the full evolution of tesseral fields in this paper, leaving it for the future. For the present, we have limited our study to the contribution of tesseral harmonics to the energy of time-latitude variations and their possible relation to the sunspot formation process. 

\begin{figure}    
\includegraphics[width=0.85\textwidth]{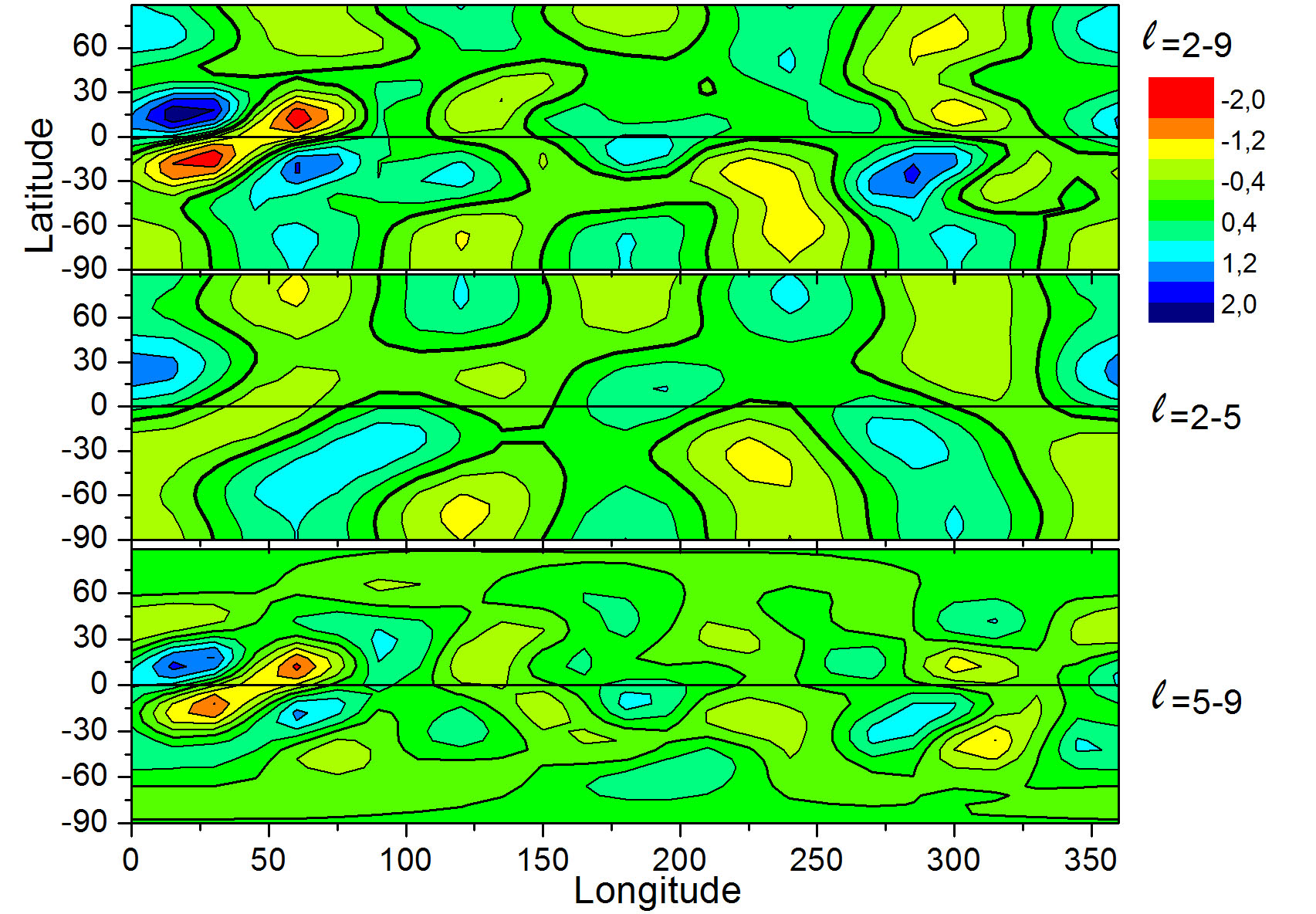}
\small
        \caption{Sample of synthesized CR1896 map taking into account harmonics $l$=2-9, $l$=2-5 and $l$=5-9       }
\label{F4}
\end{figure}

It can be expected that tesseral harmonics will be significant precisely in the equatorial zone due to the predominant influence of local fields. Figure~\ref{F4} shows by way of example  a synthesized map of the magnetic field in CR1896 (May 1995). On the upper panel, all harmonics with $l$=2-9 are taken into account, i.e., only the global dipole is excluded. One can see that the magnetic field is present both at high and at low latitudes. On the middle panel, only harmonics with $l$=2-5 were used in the synthesis. It is evident that the high-latitude regions remain almost unchanged, while all regions with a large magnetic field (active regions) in the equatorial zone have disappeared. On the third panel, only high-order harmonics ($l$=5-9) were used in the synthesis. Here, on the contrary, all structure elements at high latitudes have disappeared, and the equatorial region is clearly visible, no worse than on the upper panel.

\begin{figure}    
\includegraphics[width=0.85\textwidth]{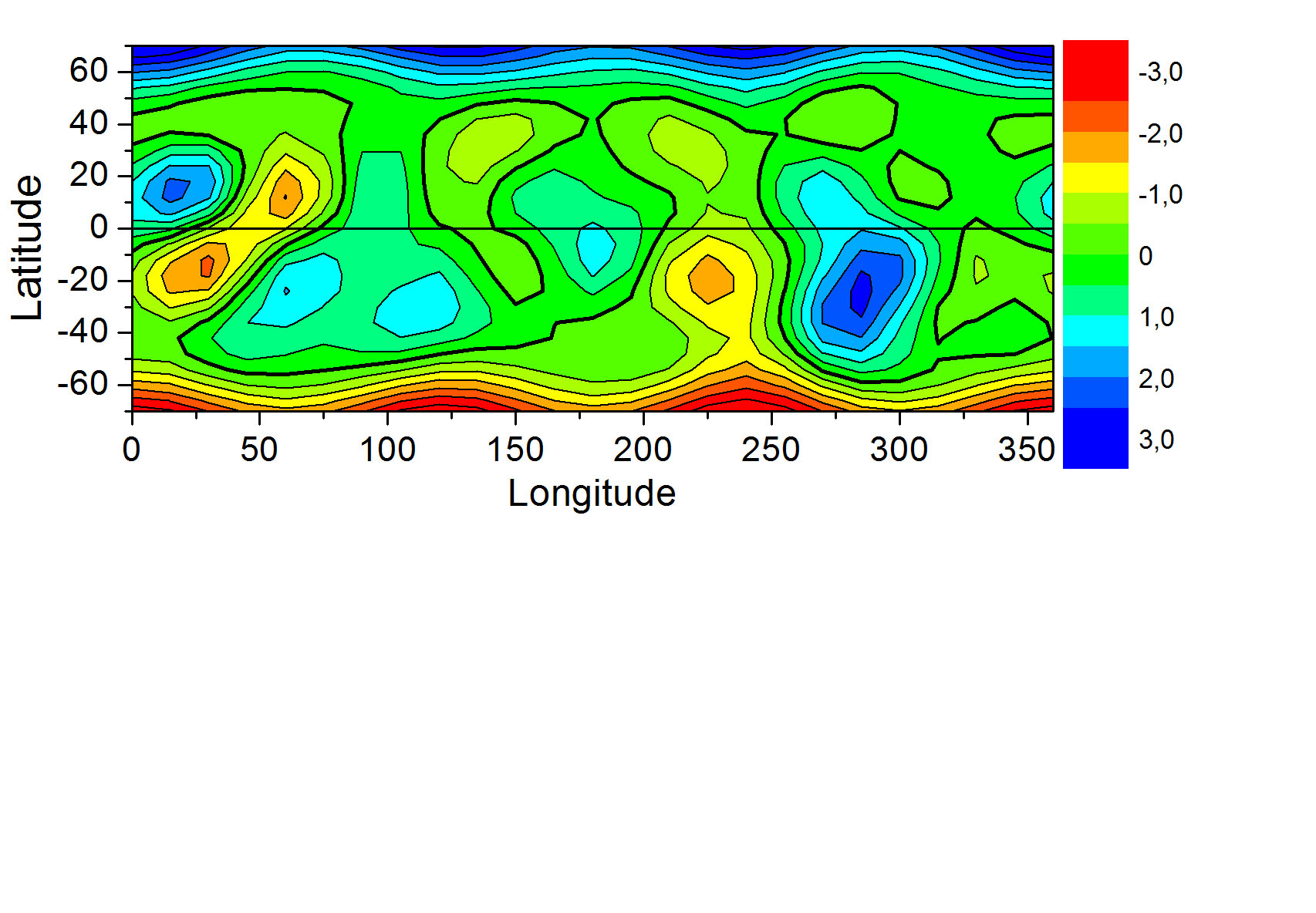}
\small
        \caption{Synthesized CR1896 map with only tesseral harmonics taken into account
       }
\label{F5}
\end{figure}

 In this case, the tesseral harmonics play a decisive role. Figure~\ref{F5} shows a synthesized map of the same rotation with only tesseral harmonics taken into account. It is clear that structural elements are visible at all latitudes, but active regions are only observed in the equatorial zone.

\begin{figure}    
\includegraphics[width=0.85\textwidth]{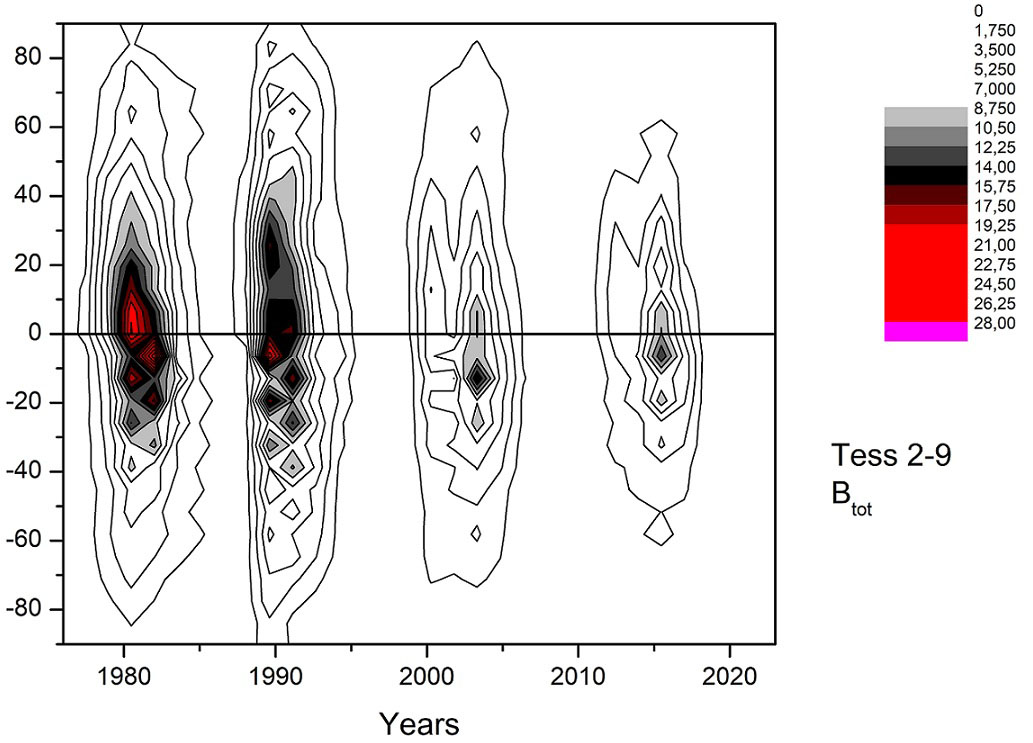}
\small
        \caption{Time-latitude diagram of the mean square magnetic field of the sum of all tesseral components.
       }
\label{F6}
\end{figure}

Indeed, Figure~\ref{F6} shows the time-latitude diagram of the mean-square magnetic field of the sum of all tesseral components (the values are given in $G^2$).
It is evident that the latter are mainly concentrated in the so-called "royal" zone at latitudes of $\pm 30^{\circ}$. The largest values given in red are located in the royal zone exclusively. Moreover, this zone gradually narrows after 1990, i.e. in cycles 22, 23, and 24. This is completely consistent with the changing height of sunspot cycles.

The equatorial zone is dominated by high-order tesseral harmonics. It is natural to expect that tesseral harmonics are associated with sunspots. A sunspot is an object, in which the magnetic field at the boundary exceeds 550 G  \citep{Obretal23b}.

\begin{figure}    
\includegraphics[width=0.85\textwidth]{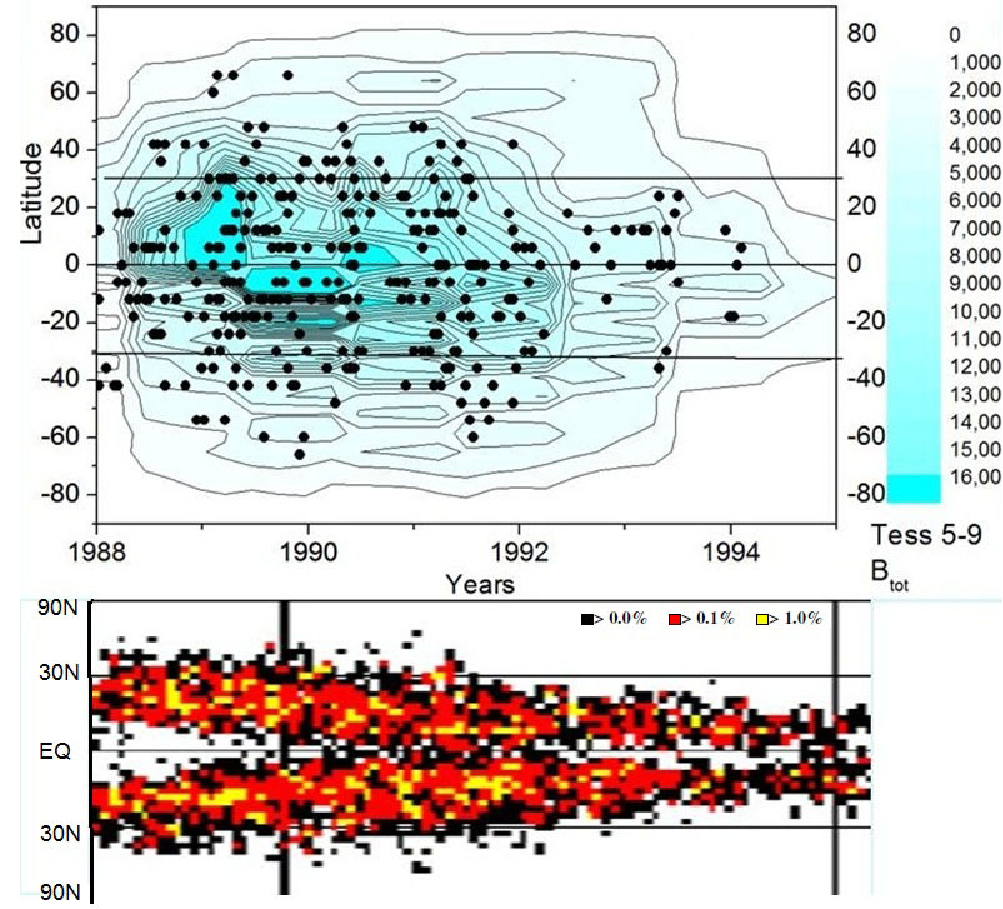}
\small
        \caption{Isolines of energy of tesseral components for 1988-1995. Black dots mark the rotations in which the magnetic field exceeded 300 nT in absolute value. The lower panel shows part of the butterfly diagram for the same period.
       }
\label{F7}
\end{figure}

Photometrically, sunspots are observed on the disk as features of low brightness, and their time-latitude diagram is known as "Maunder butterflies". Fig.~\ref{F7} compares observations of tesseral components and spots. The field lines of tesse\-ral components represent their energy for the period of 1988-1995. In this case, only the harmonics with numbers 5-9 were taken into account. Thus, the main part of the zonal and sectorial harmonics were excluded. The black dots mark rotations and latitudes, in which the magnetic field exceeded 300 nT in absolute value (The longitude was not taken into account). On WSO maps with 3 arc min resolution, this value corresponds to the field lines outlining the active regions.

The lower panel of Fig.~\ref{F7} is part of the butterfly diagram for the time interval under consideration. This is part of the diagram from Fig. 9 in \cite {hath15}. The relative area is illustrated by color on the latitude strips of equal area. There is a certain similarity between the calculated magnetic field and the butterfly diagram. On both panels, the dots are located in the $\pm30^{\circ}$ zone. The latitudinal zone of the increased magnetic field narrows gradually by the end of the cycle, as well as the zone of spots . However, there are significant differences. The butterfly diagram shows a distinct zone of avoidance at the beginning of the cycle. This means that the spot-forming activity drifts from the mid latitudes to the equator.

\begin{figure}    
\includegraphics[width=0.85\textwidth]{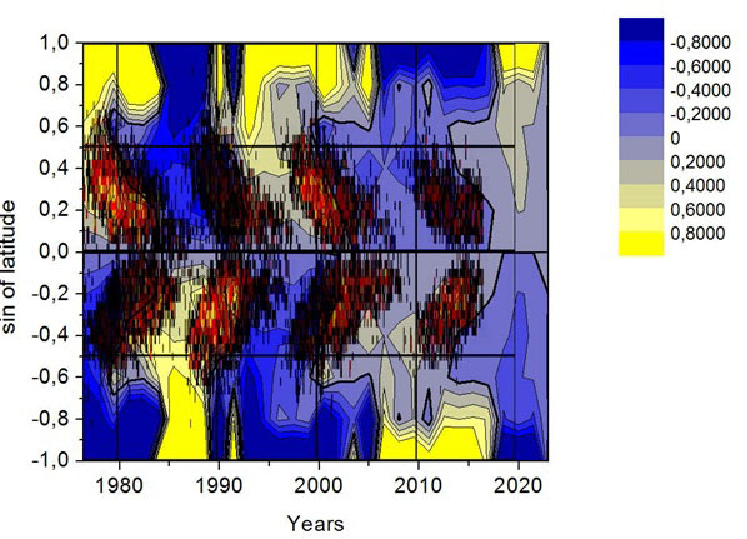}
\small
        \caption{Maunder butterflies  superimposed on the diagram of the drift of zonal harmonics.
       }
\label{F8}
\end{figure}

Figure~\ref{F8} illustrates another comparison. Here, the Maunder butterflies are superimposed on the diagram of the surface magnetic field averaged over latitude, i.e.  the field of zonal harmonics. It turns out that the butterflies are located on the boundaries of waves moving towards the equator and participate in their drift. This means that, besides a sufficiently strong magnetic field, the formation of spots requires a specific bipolar structure associated with zonal harmonics. This is not surprising, since, as a rule, the spots appear in groups, in which there are spots of different sign, and the magnetic flux is mainly closed inside the active region.

The third process of transformation of the dynamo-generated mean field into individual objects with a strong field takes place immediately beneath the photosphere. According to helioseismic studies, there is a layer between the convection zone and the surface \citep{Godro2001, Lefkos05, Lefetal07, Rozetal09}, which is often called leptocline. In this layer, the inversion of the rotation velocity occurs, as well as a change in some plasma characteristics (transparency, super-adiabaticity, and ionization degree of hydrogen and helium). The sign reversal of the rotation velocity gradient just below the photosphere was also discussed by \cite{Badobr18a, Badobr18b, Obrbad19}, who used the rotation of the coronal magnetic field as an indicator of differential rotation of the subphotospheric layers. \cite{Kitetal23} revealed the presence of a thin substructure just below the surface (Near-Surface Shear Layer, NSSL, characterized by a strong rotational gradient and self-organized meridional flows).

However, the process of transformation of the  dynamo-generated mean field  into separate objects with a strong field (i.e., spots and active regions) remains unexplored. The classical idea of bending and emergence of the subsurface toroidal field was convincingly criticized by \cite{GB19}. They showed that the scenario of early development of a bipolar magnetic region contradicts the model of emergence of an $\Omega$-shaped magnetic tube. In principle, the emergence of a magnetic field at the occurrence of an active region was confirmed by helioseismologic studies \citep{Stefetal21}, but the process of spot formation itself is still not clear. The mean field dynamo creates a toroidal mean field flow, which disintegrates into separate tubes by the action of a certain mechanism.   

\section{Results}

Let us briefly summarize our findings.

The harmonics of the mean magnetic field of all three symmetry types discussed above participate in the scenario of the nominal 11-year (or, to be more precise, 22-year) activity cycle. However, their contribution is quite different. It is reasonable to compare their behavior with the behavior of sunspots. Harmonics of each symmetry type have their own specific phase relationships with sunspots.

The shape of the solar cycle corresponding to each type of symmetry is almost independent of the cycle amplitude. The latter determines mainly the amplitude of each contribution. In this sense, the basic solar activity  cycle is similar to the eigensolution of the solar dynamo equation with amplitude modulated by some nonlinearity. 

Tesseral harmonics fill a gap in scales between the spatial distribution of the first harmonics of the mean-field  solar magnetic field  and the surface distribution of sunspots. They demonstrate properties similar to some extent to the behaviour of both, but have their own specific features as well. In this sense, taking into account tesseral harmonics makes the general shape of the solar cycle more comprehensive.

\section{Discussion}

This work, along with the set of previous ones in this series, \citep{obretal21, obretal24, obrss23}) shows that solar activity cannot be described by a single  mechanism. The solar cycle is a complex of phenomena\footnote{ Of course, there is no need to reduce all manifestation of solar magnetism just to solar cycle. Modern view on the topic can be found e.g. in \cite{SKetal23}.}. One of the most important components of this complex is the solar dynamo, understood as self-excitation of magnetic field in the form of traveling waves somewhere in the convection zone on the Sun. This activity manifests itself in various surface tracers, of which we consider the large-scale surface magnetic field. Besides the dynamo mechanism, the formation of the solar cycle involves other processes that are associated with the solar dynamo, but are not its necessary part. For example, the dynamo, as a physical process of self-excitation of  magnetic field in a moving conducting liquid, apparently works on the Earth and in spiral galaxies, but structures similar to sunspots do not form there. On the other hand, the existence and behavior of sunspots are certainly important for the operation of the solar dynamo.

This complex of physical processes is not easy to understand. We hope that it can be done by considering the components of the large-scale surface magnetic field of the Sun with different types of symmetry. The solar dynamo as an independent physical process excites a magnetic field with a certain (dipole-type) symmetry. However, the surface magnetic field on the Sun has a much more complex structure. Its emergence is obviously associated with the contribution of additional processes that have a different symmetry. By studying the behavior of the field components with a more complex structure than just a set of waves with dipole symmetry propagating from mid-latitudes to the solar equator, we hope to isolate the contributions of different physical processes to the formation of the solar cycle.

The process of formation of the eleven-year cycle seems quite clear. In the classical two-component dynamo, either a poloidal or toroidal field is set up. As one is the source of the other, both the height and the main properties of each cycle must be preserved (see Fig.~\ref{F1}). It is clear that, after the general trend is eliminated, the main components remain stable and do not change much in time. Each component has its own scenario, which is approximately the same in all cycles. However, there is a certain phase shift between the components and, after summation, a clear trend is revealed (Fig.~\ref{F2}). We can see that eleven-year cycles have a certain connection with each other, and this connection is not random. In some periods, there is a tendency to decline or to increase. The changes in the regular behavior observed over many cycles are difficult to account for by random variations in the dynamo parameters. In particular, in the  time interval under consideration, this is described fairly well by the Gleissberg quasi-hundred-year cycle. Of course, the physical process underlying this cycle deserves clarification which is far out the scope of this very paper. The trend is easily seen in Figure~\ref{F2}.

The connection between the cycles does not follow directly from the simplest dynamo model. In \citep{NP20}, the authors  considered the hypothesis of additional amplitude modulation and showed that it allows us to describe the observed characteristics of the magnetic field. The frequency modulation may also slightly change the shape of the cycle so that the cycle maximum becomes blurred and last for about two years. In addition, if this effect acts differently in different hemispheres, it leads to a mismatch of the cycle maxima in the two hemispheres, as was observed in Cycle 24, and to a pronounced asymmetry. In this case, in the growth phase of the cycle, the activity is higher in the north hemisphere (positive asymmetry) and in the decline phase, in the south hemisphere (negative asymmetry).  This, in particular, leads to the fact that the duration of the maximum phase, as well as of the polarity reversal, is about two years.

The next point is that that, in general, the mean field dynamo does not account for the appearance of activity elements such as spots. In fact, it only explains generation of the magnetic flux and its latitude  variation. Apparently, there is another mechanism that may not depend on time and on the phase of the cycle. It does not generate  new flux but transforms the mean flux into elements with a strong field.

Summarizing, we can say that embedding of various additional mechanisms participating in solar dynamo being not the obligatory part of dynamo becomes a challenge for the dynamo theory after some consensus in the basic mechanism is achieved. As an example of such mechanisms meridional circulation can be considered (see e.g. \cite{PS08, Hetal23}.

\begin{acks}
 The authors thank Todd Hoeksema for presentation of the data.
\end{acks}

\begin{authorcontribution}
All authors contributed to the study conception and design. Material preparation, data collection and analysis were performed by ASS under the guidance of VNO and DDS. The first draft of the manuscript was written by VNO and all authors commented on previous versions of the manuscript. All authors read and approved the final manuscript.
\end{authorcontribution}

\begin{fundinginformation}
No founding to declare.
\end{fundinginformation}

\begin{dataavailability}
All data on photospheric magnetic fields are available at the\\ 
http://wso.stanford.edu/synopticl.html
And Sunspot data from the World Data Center SILSO, Royal Observatory of Belgium, Brussels. Version V2. 
\end{dataavailability}


\begin{codeavailability}
\end{codeavailability}

\begin{ethics}
\begin{conflict}
The authors declare that they have no conflict of interest.
\end{conflict}
\end{ethics}
  
\bibliographystyle{spr-mp-sola}
\bibliography{sola_bibliography_example}

\end{document}